\documentclass[twocolumn,floats,aps,epsfig,nofootinbib,amssymb]{revtex4-1}
\usepackage{latexsym,graphicx,epsfig,psfrag,float,amsmath,longtable,amssymb}
\usepackage{cancel,textcomp,bm,times,graphics,hyperref}

\allowdisplaybreaks

\newcommand{\eps}{\epsilon}

\newcommand{\Nc}{N_{\text{c}}}
\newcommand{\CF}{C_{\text{F}}}

\newlength{\dslashwidth}

\def\lsim{\mathrel{\raise.3ex\hbox{$<$\kern-.75em\lower1ex\hbox{$\sim$}}}}
\def\gsim{\mathrel{\raise.3ex\hbox{$>$\kern-.75em\lower1ex\hbox{$\sim$}}}}

\catcode`@=11
\newcount\@tempcntc
\def\@citex[#1]#2{\if@filesw\immediate\write\@auxout{\string\citation{#2}}\fi
  \@tempcnta\z@\@tempcntb\m@ne\def\@citea{}\@cite{\@for\@citeb:=#2\do
    {\@ifundefined
       {b@\@citeb}{\@citeo\@tempcntb\m@ne\@citea\def\@citea{,}{\bf ?}\@warning
       {Citation `\@citeb' on page \thepage \space undefined}}%
    {\setbox\z@\hbox{\global\@tempcntc0\csname b@\@citeb\endcsname\relax}%
     \ifnum\@tempcntc=\z@ \@citeo\@tempcntb\m@ne
       \@citea\def\@citea{,}\hbox{\csname b@\@citeb\endcsname}%
     \else
      \advance\@tempcntb\@ne
      \ifnum\@tempcntb=\@tempcntc
      \else\advance\@tempcntb\m@ne\@citeo
      \@tempcnta\@tempcntc\@tempcntb\@tempcntc\fi\fi}}\@citeo}{#1}}
\def\@citeo{\ifnum\@tempcnta>\@tempcntb\else\@citea\def\@citea{,}%
  \ifnum\@tempcnta=\@tempcntb\the\@tempcnta\else
   {\advance\@tempcnta\@ne\ifnum\@tempcnta=\@tempcntb \else \def\@citea{--}\fi
    \advance\@tempcnta\m@ne\the\@tempcnta\@citea\the\@tempcntb}\fi\fi}
\catcode`@=12

\begin{document}
%\allowdisplaybreaks

\title{A first estimate of the NNLO nonresonant corrections to 
top-antitop threshold production at lepton colliders}
\author{Pedro Ruiz-Femen\'ia}
\affiliation{Instituto de F\'isica Corpuscular (IFIC),
CSIC-Universitat de Val\`encia, E-46980 Paterna, Spain}
%\preprint{IFIC/14-XX}
\date{\today}

\begin{abstract}
\noindent
We compute the dominant term
in the expansion in $\rho=1-M_W/m_t$ of 
the unknown next-to-next-to-leading order (NNLO) nonresonant
contributions to the $e^+e^-\to W^+W^- b\bar{b}$ total cross section
at energies close to the top-antitop threshold. Our analytic
result disagrees with a previous calculation by other authors~\cite{Penin:2011gg}.
We show that our determination has the correct infrared structure needed to cancel the 
divergences proportional to the top width arising in the resonant
production of the same final state,
and we point out to a missing contribution in the computation
of~\cite{Penin:2011gg} to explain the discrepancy.  
\end{abstract}

\maketitle
%%%%%%%%%%%%%%%%%%%%
\section{Introduction}
%%%%%%%%%%%%%%%%%%%%

Stringent tests of the electroweak symmetry breaking sector of the Standard Model 
shall require a very precise knowledge of the top-quark properties, and most crucially of its mass.
The measurement of the top-antitop production cross section line shape at a future
$e^+e^-$ collider could allow us to determine the top mass with an uncertainty
below 100~MeV~\cite{Martinez:2002st}
which is
beyond the reach of hadronic colliders.
Substantial effort has been put in the past to increase the accuracy
in the calculation of the $e^+e^-\rightarrow t\bar{t}$ cross section
with the use of perturbative methods that include all-order resummations of 
QCD corrections in the threshold region. The latter arise from Coulomb-gluon 
exchange between the top and the antitop that are produced nearly on-shell 
(resonantly), and are responsible for
the characteristic $t\bar{t}$ threshold line shape, where a remnant of a toponium
1S resonance, smoothed out by the large top width ($\Gamma_t\approx 1.5$~ GeV), is
visible. For a long time, the aim was put in the calculation of the resonant QCD 
corrections with the help of the non-relativistic effective field theory 
of QCD (NRQCD): within the power counting defined by $v\sim \alpha_s$, with
$v$ the relative velocity of the top and antitop, 
next-to-next-to-leading logarithmic (NNLL) and fixed-order
N${}^3$LO results have become available (for the most recent calculations
concerning the resonant pieces see~\cite{Hoang:2013uda,Beneke:2013jia,Beneke:2013kia}).

The high accuracy reached in the evaluation of QCD corrections has also
called for an assessment of the electroweak corrections and higher-order
effects related to the top-quark instability. In the resonant side, the top decay is 
simply accounted for at leading order by including the top width in the NRQCD
top-antitop propagator, $(E-\mathbf{p}^2/m_t +i \Gamma_t)^{-1}$, which enforces 
the counting $\Gamma_t\sim m_t v^2 \sim m_t \alpha_s^2$, or equivalently,
$\alpha_{\mbox{\tiny EW}}\sim \alpha_s^2$. Beyond leading order,
apart from subleading resonant electroweak effects~\cite{Hoang:2004tg},
we have to consider the possibility that the physical final state,
$W^+W^- b\bar{b}$ (treating $W$ bosons as stable)
is produced non-resonantly, {\it i.e.} by processes that do not
involve a nearly on-shell $t\bar{t}$ pair. The leading nonresonant
effects are NLO for the cross section and account for
the full-theory contributions where one of the $bW$ pairs 
is produced from an on-shell top, while the other pair emerges
from a highly virtual top or directly without an intermediate top.
They were calculated in~\cite{Beneke:2010mp}, and shown to yield a constant
negative shift of order 3\% with respect the LO (resonant) result above threshold
and to become particularly relevant below, where the resonant contributions rapidly vanish. 
At NNLO, the nonresonant corrections arise from attaching real and virtual
gluons to the NLO diagrams. They are only known when 
cuts on the invariant masses 
on the $bW$ pairs of size $\Lambda^2$, with $m_t\Gamma_t\ll\Lambda^2\ll m_t^2$,
are applied~\cite{Hoang:2010gu,Jantzen:2013gpa}. The extrapolation
of the results of the latter works for the total cross section case,
$\Lambda^2_{\rm{max}}=m_t^2-M_W^2$, suggests that the NNLO contributions are only 
one half smaller than the NLO ones.
Given that the simulation studies of a $t\bar{t}$ threshold scan at a future $e^+e^-$ collider
are able to identify top pair events with high purity without applying cuts on the top and antitop 
invariant masses, a calculation of the NNLO nonresonant
effects for the totally inclusive cross section will become necessary
in order to reach a percent accuracy in the theory input.

In this note we provide an approximation to the  NNLO nonresonant
total cross section by performing an expansion in
the parameter $\rho=1-M_W/m_t$. This expansion was applied in~\cite{Penin:2011gg}
to determine the NLO nonresonant contributions up to very high orders in $\rho$, confirming
the results from~\cite{Beneke:2010mp}.
Despite the fact that $\rho$ is not a small
parameter at its physical value ($\rho\approx 0.5$), it was found in~\cite{Penin:2011gg} 
that the dominant term in this expansion
gives a result that differs from the exact one by only 5\%. 
Ref.~\cite{Penin:2011gg} also provided the 
leading and subleading pieces in $\rho$ for the NNLO corrections, scaling as $\rho^{-1}$ and 
$\rho^{-1/2}$, respectively. 
However, as pointed out in~\cite{Jantzen:2013gpa}, the infrared 
divergent part of the NNLO $\rho^{-1}$ term given in~\cite{Penin:2011gg} does not match the structure needed 
to cancel the well-known finite-width divergences 
$\propto \frac{\alpha_s \Gamma_t}{\epsilon}$ that arise in the NNLO
resonant cross section at the same order in $\rho$. 
Such cancellation is mandatory since the full-theory calculation for the same
process is finite~\cite{Beneke:2008cr} (see also~\cite{RuizFemenia:2012ma}).
This mismatch is corrected in this work by providing a new evaluation of the 
$\rho^{-1}$ term which yields the correct infrared structure, and
that can be combined with the NNLO
resonant corrections in a regularization-scheme independent way.

%%%%%%%%%%%%%%%%%%%%%%%%%%%%%%%%%%%%%%%%%%%%%%%%%%%%%%%%%%%%%%%%%%%%%%%%%%%%$
\section{NNLO nonresonant cross section: expansion in $\rho=1-M_W/m_t$}
%%%%%%%%%%%%%%%%%%%%%%%%%%%%%%%%%%%%%%%%%%%%%%%%%%%%%%%%%%%%%%%%%%%%%%%%%%%%%
\label{sec:EFTdivs}

\begin{figure}[t]
\begin{center}
\includegraphics[width=3.cm]{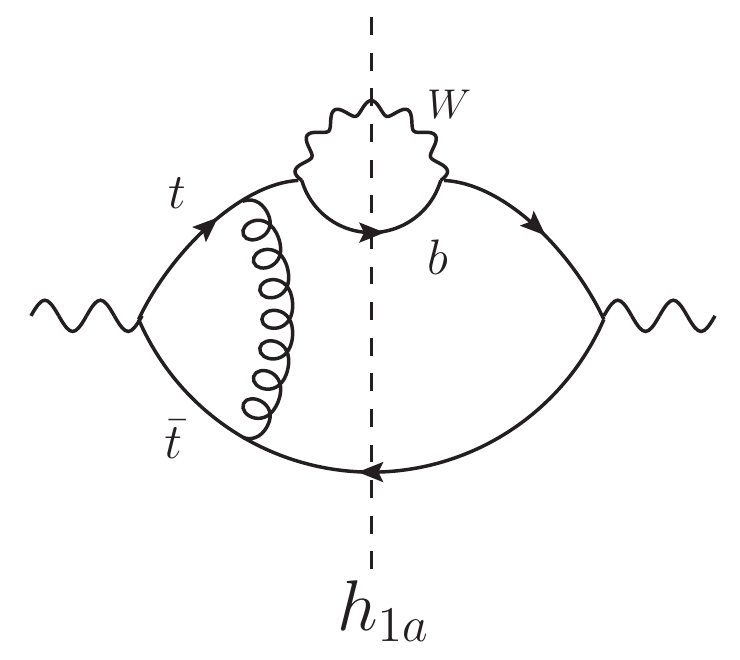}
\hspace*{0.2cm}
\includegraphics[width=3.cm]{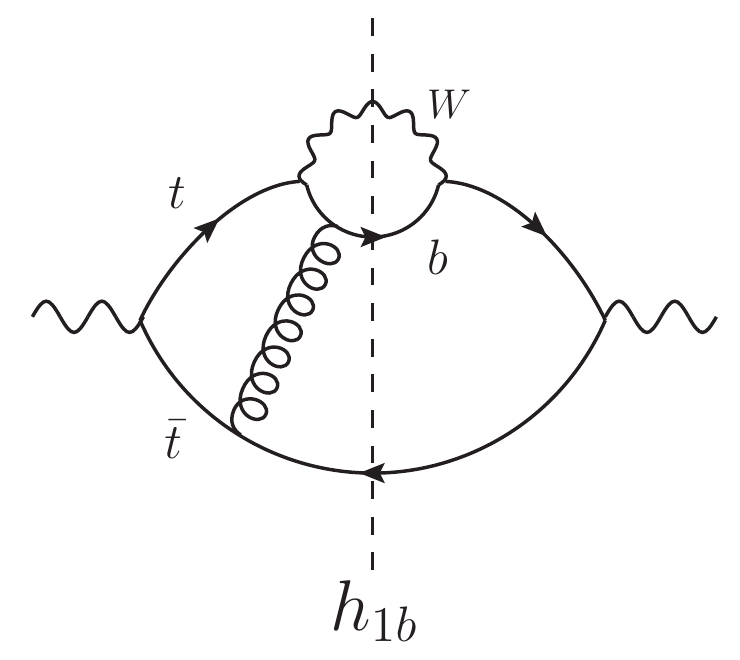}
\hspace*{0.cm}

\caption{${\cal O}(\alpha_s)$ corrections to the two-loop forward-scattering  diagrams 
which provide the ${\cal O}(\rho^{-1})$ NNLO nonresonant approximation. The $e^+e^-$ external legs have not been drawn.
The symmetric diagrams with the gluon in the r.h.s of the cut, or with $\bar{b} W^- t$ cuts
are not shown.}
\label{fig1}
\end{center}
\end{figure}

The total $e^+e^-\to W^+W^- b\bar{b}$ cross section  is 
conveniently obtained by extracting
the cuts of the $e^+e^-$ forward-scattering amplitude related to this final state.
The separation of resonant and nonresonant effects 
can be achieved systematically
by expanding the full electroweak theory diagrams of the $e^+e^-$ forward-scattering
amplitude according to regions with small 
($p_t^2-m_t^2\sim m_t\Gamma_t$) and large ($p_t^2-m_t^2\sim m_t^2$) virtuality 
in the top and antitop lines~\cite{Beneke:2007zg,Beneke:2010mp}. The leading nonresonant contributions 
are given by two-loop diagrams of order 
$\alpha_{\mbox{\tiny EW}}^3$ with 3-particle cuts $bW^+\bar{t}$ and $\bar{b}W^-t$,
and are suppressed by $\alpha_{\mbox{\tiny EW}}/v\sim v$ (NLO) with respect the leading 
order (resonant) cross section. The complete set of diagrams can be found in~\cite{Beneke:2010mp}. 
The EFT power-counting $\alpha_s\sim\alpha_{\mbox{\tiny EW}}^{1/2}$ implies that the NNLO
nonresonant corrections arise from (virtual and real) QCD corrections to the NLO diagrams.
The number of diagrams at this order exceeds 100, but only the two shown in 
Fig.~\ref{fig1} contribute to the dominant term in the $\rho$-expansion, as we explain next.

\begin{figure}[t]
\vspace*{0.1cm}
\begin{center}
\includegraphics[width=7.8cm]{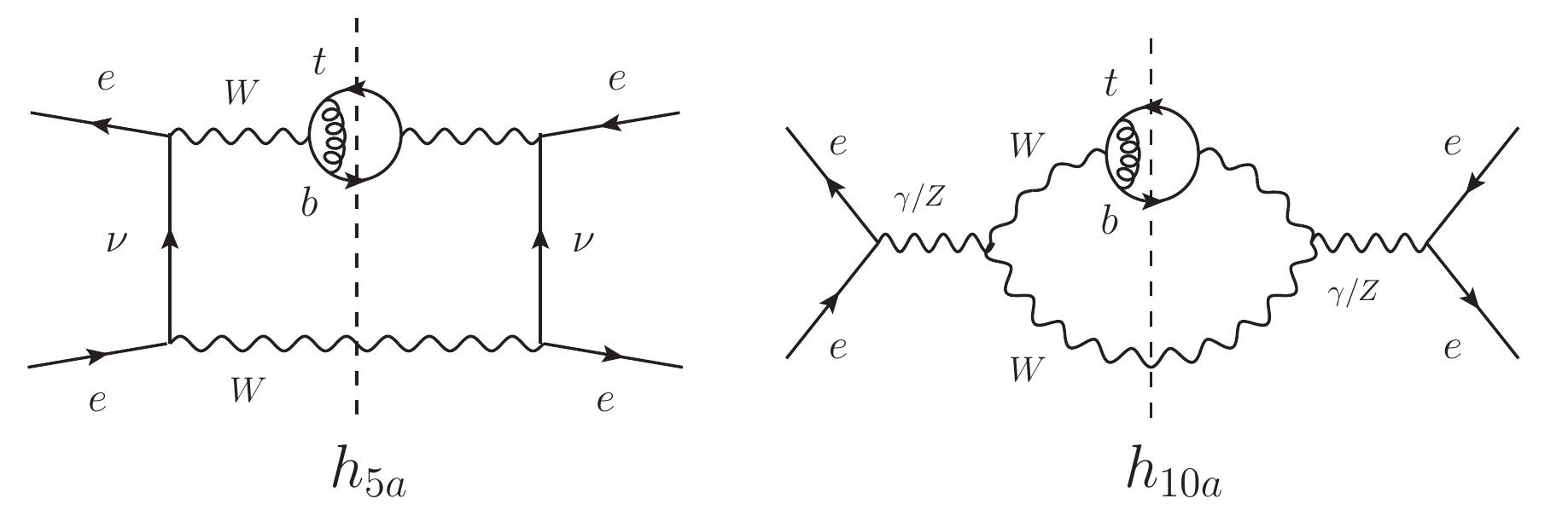}
\caption{Example diagrams contributing to the NNLO nonresonant cross section
with virtual $W$-bosons.}
\label{fig2}
\end{center}
\end{figure}

After integrating over other phase-space variables,
the NNLO nonresonant virtual contributions to the total cross section can be casted in the form~\cite{Jantzen:2013gpa}
\begin{equation}
\int^{1}_{x} \frac{dt}{(1-t)^{r+ n \epsilon}} \, h(t,x) \,,
\label{eq:genform}
\end{equation}
where the outer integration variable, $t\equiv p_t^2/m_t^2$, corresponds to the 
invariant mass of the $bW^+$ subsystem, $p_t=p_b+p_W$, and 
$x\equiv m_t^2/M_W^2$. The functions $h(t,x)$ contain the
result of gluon loop integrals like those of Figs.~\ref{fig1},\ref{fig2}, 
whereas the explicit powers of $(1-t)$ arise from top propagators and
phase-space factors (see~\cite{Jantzen:2013gpa} for details). 
Since $1-x=\rho(2-\rho)$, the expansion 
for small $\rho$ can be traded by an expansion in $(1-x)$. Let us first note that the integration domain 
in (\ref{eq:genform}) implies that $0\le 1-t\le 1-x$. Therefore, 
the leading
terms in $(1-x)$ 
can be identified by asymptotically
expanding  the integrand $h(t,x)$  around
the phase-space endpoint $t=1$, {\it i.e.} in powers of $(1-t)$,
which are transformed into powers of $(1-x)$ upon integration over $t$.
The expansion of the NNLO contributions in the 
endpoint was carried out in~\cite{Jantzen:2013gpa} up
to terms of order $(\Lambda/m_t)^0\ln\Lambda$, where the lower limit
in the $t$-integration was limited by $(1-\Lambda^2/m_t^2)$
to regulate the endpoint behaviour. The straightforward 
replacement $\Lambda^2/m_t^2\to 1-x$
into the results  of~\cite{Jantzen:2013gpa} yields that the
leading terms scale as $(1-x)^{-1}$ and arise only from diagrams $h_{1a}$ and $h_{1b}$.
Some caution has to be taken with the diagrams involving intermediate $W$-bosons, since
in the limit $\rho\to 0$ where $M_W\to m_t$ these become nonrelativistic and
their propagator contributes  with a factor 
$1/(m_t^2-M_W^2)\sim 1/m_t^2(1-x)$ to the amplitude.
The NNLO diagrams where this enhancement is most relevant are those of Fig.~2. Taking
into account that its endpoint behaviour is $(\Lambda/m_t)^3\sim (1-x)^{3/2}$~\cite{Beneke:2010mp},
an overall
scaling  $(1-x)^{-1/2}\simeq \rho^{-1/2}$  is then found for diagram $h_{5a}$, in agreement with the 
findings of~\cite{Penin:2011gg}. Diagram $h_{10a}$ carries a further $(1-x)^2$
suppression factor because $s$-channel $W^+W^-$  production is suppressed in the
non-relativistic limit. 
We should also note that the endpoint limit taken in~\cite{Jantzen:2013gpa} 
assumed that
$(1-x)\sim {\cal O}((1-t)^0)$ for 
the asymptotic expansion in powers of $(1-t)$
of diagrams where the bottom propagator\footnote{The components of the (massless) bottom quark momentum 
read $p_b^0=|\vec{p}_b|=m_t(t-x)/2\sim m_t(1-x)/2$ in the top rest frame, where $p_t=(p_t^0,\mathbf{0})$.},
$1/(p_b+k)^{2}$, is part of the gluon loop with momentum $k$, such as in $h_{1b}$. 
For small $(1-x)$, however, 
we have to consider that $(1-t)$ but also $(1-x)$ are small quantities in
the expansion of the bottom propagator, which 
invalidates the naive use of~\cite{Jantzen:2013gpa} to infer
the scaling in $(1-x)$ for those cases. A careful examination of the 
regions involved in diagrams with a bottom quark propagator in the loop
reveals that this caveat applies only to $h_{1b}$. Nevertheless,
the $\rho^{-1}$ scaling found for
$h_{1b}$ with the naive procedure turns out to be correct, as 
shown below by
explicit computation. 
%
%%%%%%%%%%%%%%%%%%%%%%%%%%%%%%%%%%%%%%%%%%%%%%%%%%%%%%%%%%%%%%%%%%%%%%%%%%%%$
\section{Results}
%%%%%%%%%%%%%%%%%%%%%%%%%%%%%%%%%%%%%%%%%%%%%%%%%%%%%%%%%%%%%%%%%%%%%%%%%%%%%
\label{sec:results}
For consistency with the calculation of the resonant contribution to the cross section,
where finite-width divergences are regulated dimensionally,
we deal with the infrared endpoint singularities of the nonresonant
side using the $\overline{\rm MS}$ regularization scheme in $d=4-2\epsilon$. 
The $\rho^{-1}$ terms from diagrams $h_{1a}$ and $h_{1b}$ arise from
potential gluon momentum $k^0\sim\mathbf{k}^2/m_t\sim m_t(1-t)$ running through the QCD loop. 
In the case of $h_{1a}$, the leading contribution from the potential region 
gives a $(1-t)^{-1/2-\epsilon}$ term,
while the two top propagators plus the phase-space factor provide an extra factor
$(1-t)^{-3/2-\epsilon}$. A dependence on $x$ only enters through the off-shell top width
\begin{align}
\Gamma_t(t) =  &\ \Gamma_t^{\rm Born} \, \frac{(4\pi)^\epsilon \, \Gamma(1-\epsilon)}{\Gamma(2-2\epsilon)}\,
  \,m_t^{-2\epsilon}\,t^{-1+\epsilon}\,
\nonumber\\[-1mm]
& \, \times \frac{1+2(1-\epsilon){x\over t}}{1+2x}\,
\frac{(t- x)^{2-2\epsilon}}{(1-x)^2}
\,.
\label{eq:Gammatoff}
\end{align}
with $\Gamma_t^{\rm Born}$ the tree-level top decay width.
Upon integration over $t$, powers of $(1-t)$ and $(t-x)$ are converted into 
powers of $(1-x)$, and  
the outcome is proportional to $\Gamma_t^{\rm Born}/(1-x)$, see~(\ref{eq:H1a}) below.
The computation of diagram $h_{1b}$ is more involved, because after expanding
the propagators in the loop according to the potential scaling, it still depends on the antitop and bottom quark
three-momenta. Using a Mellin-Barnes representation to perform the $d^dk$ integral 
we get a result for the integrand in (\ref{eq:genform}) whose pole structure reveals
that, unlike the case of $h_{1a}$, the leading-order term in the asymptotic expansion in
the small parameter $(1-x)$ for the $t$-integration arises from the region
$(1-t)\sim(1-x)^2$~\cite{BJantzen}.

In this way, the contribution of the two diagrams to the hadronic tensor $H$ of the $e^+e^-$ 
forward-scattering amplitude, as defined in~\cite{Jantzen:2013gpa}, reads
\begin{widetext}
\vspace*{-0.3cm}
\begin{align}
  \label{eq:H1a}
  H_{1a}
   = - 2 \,N_\eps \, v_t^\text{L} v_t^\text{R}  \,(1-x)^{-1-4\epsilon}\,
     \frac{2^{2\epsilon}  \,e^{3\epsilon \gamma_E}}{\pi}  \,
    \frac{\Gamma(2-\epsilon) \,\Gamma(1/2+\epsilon) \,\Gamma(1/2-\epsilon)}{\epsilon\,(1-4\epsilon^2)\,\Gamma(2-4\epsilon)} \,,
\end{align}
\vspace*{-0.2cm}
\begin{align}
  \label{eq:H1b}
  H_{1b}
   = \,i^{4\epsilon} N_\eps \, v_t^\text{L} v_t^\text{R}  \,(1-x)^{-1-6\epsilon}\,
     \frac{4\,e^{3\epsilon \gamma_E} \,(3-2\epsilon)}{3\,(1+4\epsilon)\,(1-6\epsilon)}
    \, 
    \frac{\Gamma(4\epsilon)\,\Gamma(2-\epsilon) \,\Gamma(1/2+\epsilon) \,\Gamma^2(1/2-\epsilon)\,\Gamma(1-4\epsilon)}
         {\sqrt{\pi}\,\Gamma(1-2\epsilon)\,\Gamma(2-2\epsilon)\,\Gamma(1/2+2\epsilon)\,\Gamma(1/2-3\epsilon)} \,,
    \,
\end{align}
\end{widetext}
where $N_{\eps}  = m_t \Gamma_t^{\rm Born} \, \Nc \CF \,
    \frac{\alpha_s}{4\pi}(\mu^2/m_t^2)^{3\eps}$
and $\mu$ is the scale introduced in dimensional regularization. $v_t^\text{L,R}$
are the vector 
couplings to the photon and $Z$-boson of the top quarks at the left-hand and right-hand 
of the diagram, following the conventions of~\cite{Jantzen:2013gpa}. We note that both $H_{1a}$ and $H_{1b}$ produce $1/\epsilon$
divergences, which originate at the endpoint. Using the  relation between the  hadronic tensor contributions
and the cross section given in~\cite{Jantzen:2013gpa}, we  find for the dominant term in the expansion
in $\rho$ of the NNLO nonresonant cross section the result
\begin{widetext}
\vspace*{-0.3cm}
\begin{align}
  \label{eq:sigmanonres}
  \sigma_{\rm non-res}^{(2),\rho}
    = & \,  \frac{8\pi \alpha^2}{s} \,
      m_t\,\Gamma_t^{\rm Born} \, \Nc     
       \CF \, \alpha_s \, 
      \biggl[ \, \frac{Q_t^2 }{s} 
        - \,\frac{2 \,Q_t v_t v_e }{(s-M_Z^2)}  +  \, \frac{v_t^2(v_e^2 + a_e^2) \, s}{(s-M_Z^2)^2} \, \biggr]
  \nonumber\\[-1mm]
 &
     \,\times \frac{1}{\rho}\,\biggl( \frac{1}{2\epsilon}+\frac{2}{3}-\ln {\rho\over 2}+ \ln \frac{\mu_{\rm soft}^2}{m_t^2} 
  + {\cal O}(\rho^{1/2})  \biggr)
  \,,  
\end{align}
\end{widetext}
where we have replaced $\mu^{6\eps} \to \mu_{\rm hard}^{2\eps} \,\mu_{\rm soft}^{4\eps}$
and taken $\mu_{\rm hard}=m_t$ because one integration is associated with the hard
decay $t\to bW^+$. 
The $1/\epsilon$ part of (\ref{eq:sigmanonres}) can be shown to cancel 
against the finite-width divergence $\alpha_s\Gamma_t/\epsilon\times 1/\rho$
in the resonant NNLO cross section, that arises from the insertion of the
absorptive matching coefficients $C_p^{(v/a),\rm abs}$, see Eq.~(10) of \cite{Jantzen:2013gpa}.
Since the latter coefficients were computed at $d=4$~\cite{Hoang:2004tg}, for a
consistent addition of resonant and nonresonant finite pieces, the three-loop
resonant diagrams producing these finite-width divergences
should also be evaluated in $d$ dimensions\footnote{Alternatively, a subtraction scheme
that regulates simultaneously the divergences in the resonant and nonresonant parts of the
relevant 
diagrams could be introduced. That analysis is beyond the scope of this note and is postponed
to a future publication}.

\begin{figure}[t]
\hspace*{-0.5cm}
\includegraphics[width=7.5cm]{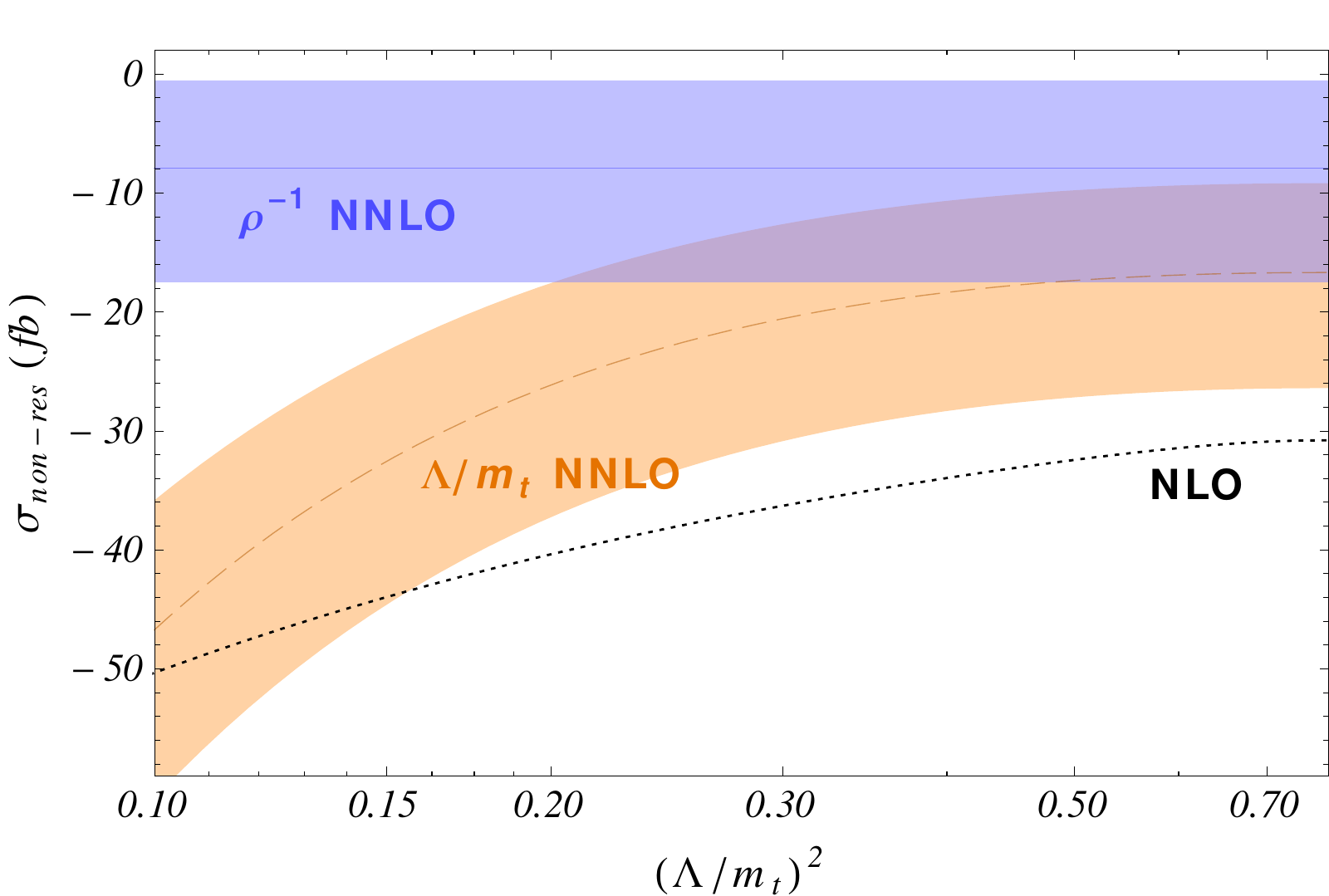}
\caption{NNLO non-resonant contribution to 
$t\bar{t}$ production computed at $s=4m_t^2$: 
$1/\rho$ approximation to the total cross section (\ref{eq:sigmanonres}) (solid blue line)
and first three terms in the $\Lambda/m_t$ expansion from \cite{Jantzen:2013gpa}
(dashed orange line), dropping the $1/\epsilon$ and using 
$\alpha_s\equiv \alpha_s(\mu_{\rm soft})$ with $\mu_{\rm soft}=30~{\rm GeV}$.
The bands show the result of varying $\mu_{\rm soft}$ in 
the interval 15-60~GeV. 
For comparison, the NLO non-resonant contribution
(dotted black line), $\sigma_{\rm non-res}^{(1)}$ from \cite{Beneke:2010mp},
is also shown. } 
\label{fig3}
\end{figure}

The size of the NNLO estimate (\ref{eq:sigmanonres}) (once the $1/\epsilon$
divergence is removed) is compared to the (exact) NLO result and to the endpoint
NNLO approximation with the 
first three terms in $(\Lambda/m_t)$ included in Fig.~\ref{fig3}. 
The dependence on $\mu_{\rm soft}$  shown in Fig.~\ref{fig3} is compensated 
by a similar logarithmic dependence associated to the canceled finite-width divergences in the resonant side. 
For $\mu_{\rm soft}=30~{\rm GeV}$ our estimate for the NNLO non-resonant corrections to the total cross section yields
$\sigma_{\rm non-res}^{(2),\rho}\simeq -8$~fb, representing a $\lsim 1\%$ negative shift 
with respect the LO resonant result above the threshold. 
We note that the extrapolation of the endpoint NNLO approximation for values of $\Lambda$ close to $\Lambda_{\rm{max}}$
approaches the estimate for the total cross section given in this work, although it overestimates
its value roughly by a factor of two. 
The validity of the $\rho$-expansion at NNLO should be further assessed  by
computing the next-to-leading order term, of ${\cal O}(\rho^{-1/2})$,
that gets contributions from a number of sources, some of which have been identified
by~\cite{Penin:2011gg}. 

Let us finally comment on the result for $\sigma_{\rm non-res}^{(2),\rho}$
obtained in the latter reference. It is argued in~\cite{Penin:2011gg} that the nonresonant
diagram $h_{1b}$ does not contribute in their framework 
because it was found to vanish in~\cite{Melnikov:1993np}. The computation
of~\cite{Melnikov:1993np} corresponds to a situation where the (potential) gluon momentum $|\mathbf{k}|$ is neglected
with respect $m_t-M_W=\rho\, m_t$ in the phase-space integration of the $t\to bWg$ subgraph. In the $\rho\to0$ limit this simplification
is not longer allowed, as we have seen that the dominant term in $\rho$ actually comes from gluon momentum
with $|\mathbf{k}|/m_t\sim (1-x)\simeq 2\rho$. To agree with our findings, $h_{1b}$ should also yield a 
contribution the $\rho\to0$ limit in the scheme of~\cite{Penin:2011gg}. Diagram $h_{1b}$ might also
modify the next-to-leading result in $\rho$ given in~\cite{Penin:2011gg}, since
it potentially produces sub-dominant terms.

\noindent
\subsubsection*{Acknowledgments}
I thank B.~Jantzen for helpful discussions.
This work has been supported in part by the Spanish
Government and ERDF funds from the EU Commission
[Grants No. FPA2011-23778, No. CSD2007-00042
(Consolider Project CPAN)] and by Generalitat
Valenciana under Grant No. PROMETEOII/2013/007.

%%%%%%%%%%%%%%%%%%%%%%%%%%%%%%%%%%%%%%%%%%%%%%%%%%%%%%%%%%%%%%%%%%


\begin{thebibliography}{99}
\bibitem{BJantzen}
  This finding has been cross-checked using a {\sc Mathematica} code to extract poles from Mellin-Barnes integrals written by B.~Jantzen.
\end{thebibliography}
\end{document}